\documentclass[10pt,a4paper,english]{article}
\usepackage{setspace}
\usepackage[german,english]{babel}
\usepackage[T1]{fontenc} 
\usepackage[latin1]{inputenc}
\usepackage{amsfonts}
\usepackage{amsmath}
\usepackage{latexsym}
\usepackage{amssymb}
\usepackage{epsfig}
\usepackage{moreverb}
\usepackage{rotating}
\usepackage{enumerate}
\usepackage{graphics, graphicx,wrapfig}
\usepackage{graphicx,xcolor}
\usepackage{fancybox}
\usepackage{picinpar,varioref,floatflt}
\usepackage{ae}
\usepackage{longtable}
\usepackage{textcomp}
\usepackage{float}
\usepackage{url}
\usepackage{unizhdt}
\usepackage{listings}
\usepackage{multirow}
\usepackage[bottom]{footmisc}
\usepackage{titlesec}
\usepackage{tikz}
\usepackage{color, colortbl}

\lstdefinelanguage{mandala}{
  	sensitive=true, 
	numbers=left,
 	basicstyle=\small,
     stepnumber=1,
	keywords = [1]{module, type, val, capability, import},
  	keywordstyle=[1]\color{orange},
	keywords = [2]{public, private, open, protected},
  	keywordstyle=[2]\color{blue},
	keywords = [3]{risk, default, ref, borrowed},
  	keywordstyle=[3]\color{violet},
	keywords = [4]{pure, init, dependent, active},
  	keywordstyle=[4]\color{cyan},
	keywords = [5]{Persist, Copy, Drop, Withdraw, Master, Modify},
  	keywordstyle=[5]\color{olive},
	keywords = [6]{return, modify, with, case, of, =>},
  	keywordstyle=[6]\color{teal},
}

\begin{document}
\selectlanguage{english}

\author{Markus Knecht\\[0.5cm]{\small Supervisor: Prof. Dr. Burkhard Stiller}}

\title{Mandala: A Smart Contract Programming Language}




\maketitle


\onehalfspacing
\pagenumbering{arabic}

\begin{abstract} 
Smart contracts on a blockchain behave precisely as specified by their code. A vulnerability in this code can lead to unexpected behaviour, which is hard to fix because a blockchain does not allow to change smart contract code after its deployment.
\par
Such vulnerabilities have led to several incidents. In the aftermath of such an event, a hard-fork between Ethereum and Ethereum classic was the result.  This thesis proposes to develop a new smart contract programming language with the primary focus on safety, auditability, and the intention to prevent as many of the known categories of vulnerabilities by design as possible. The programming language's code is validated during deployment and afterwards isolated from other smart contracts running on the same blockchain to enforce compile-time guarantees during runtime. The designed programming language does evaluate new concepts and paradigms rarely used in non-smart contract environments for their potential benefit in a smart contract environment.
\end{abstract}
\section{Introduction and Motivation}\label{sec:intro}
Smart contracts are autonomous entities managing valuable assets, such as cryptocurrency coins or ownership certificates. The code of a smart contract entirely defines its behaviour. A flaw in the code can result in the loss or theft of the controlled assets. Developing bug-free software is challenging even for skilled professionals~\cite{Rai:2014}. Smart contract blockchains do not allow to revert the execution of a transaction. Because of the value at stake, flaws and vulnerabilities in deployed smart contracts have a heightened impact compared to flaws and vulnerabilities contained in the code of non-smart contract applications (classical applications). 
\par
This research proposal presents an approach to improve the current situation for developers and auditors alike by proposing a new smart contract programming language called \textit{Mandala} that has a focus on preventing such flaws and vulnerabilities. This is achieved by evaluating new concepts and paradigms rarely used in non-smart contract environments for their potential benefit in a smart contract environment and incorporating the results into \textit{Mandala}.

\subsection{Prior Work}
During the Fast-Track Master proceeding this thesis, a paper with the title \textit{SmartDEMAP: A Smart Contract Deployment and Management Platform}\cite{Knecht:2017} was published. The paper analysed the trade-off between smart contracts that can not be changed after deployment (immutable, trustless) and the use of patterns that allow exchanging the logic of a smart contract application after deployment (mutable, trust needed). The paper proposed a solution in between where on chain management platform guards the process of exchanging the logic of a smart contract application with the goal of reducing the trust needed while still allow bugs to be fixed and new features to be added. It further proposed the idea that such a management platform could benefit from a custom smart contract programming language that is aware of the capabilities of the platform and can use them to their full extent. 
 
\subsection{Mandala Overview}\label{sec:overview}
\textit{Mandala} is a statically typed language that has a type system with algebraic datatypes at its core. \textit{Mandala} is highly predictable and achieves that by only providing function calls where the called function implementation is known during compilation (static method dispatch). This property is strengthened further by not allowing recursion and only providing loops where the maximal number of iterations is known during compilation. As a result, \textit{Mandala} is not Turing complete but in return allows to calculate during compilation an upper bound for the resources consumed when executed. 
\par
\textit{Mandala} provides a novel approach for application isolation,  access control to resources as well as ownership tracking. It does this by recording and tracking capabilities on the types of values, and only with the respective capability someone can create, inspect, drop or copy values of a certain type. This concept is generic and allows developers to define custom capabilities which then allow safeguarding function calls against callers that are not in possession of a value with the required capability. This makes it possible to check access control and express ownership transfers (for example of an asset) statically at compiletime instead of dynamically at runtime. To prevent shared state problems \textit{Mandala} does provide a minimalistic effect system and for error handling \textit{Mandala} uses a concept somewhere in between error codes and exceptions.

\subsection{Motivation}\label{sec:motiv}
Smart contracts and blockchains are a relatively new technology compared to other programmable platforms like cloud systems. They promise to deliver a platform where applications can be run, which can not be stopped, corrupted or censored and work across borders and at the same time deliver an unprecedented amount of transparency. For the first time applications that do not require any trust in a third party to guarantee the correct and fair execution of an application becomes feasible in practice. With blockchains and smart contracts, everybody can inspect the code that runs on it at any time and can be ensured that exactly this code is executed and nothing else.  Blockchains have the potential to be a disruptive technology with a significant impact on computer sciences and enable whole new industries. This is the motivation behind choosing smart contracts and blockchains as the area to advance through scientific research with this research proposal.
\par
When developing classical applications, there is always the possibility to redeploy software to fix bugs if they appear during productive use. If a bug leads to damage, most of the time there is an official party that can repair or minimise the damage by deploying new code and changing entries in a database or other state holding systems. A smart contract, on the other hand, thrives on the fact that its code is immutable and state transitions can only occur according to the rules described by its code and once executed are irreversible. Because of this, the capability to fix problems with the classical approach does not exist in a smart contract environment. Flaws in a deployed smart contract can cause permanent and unrecoverable damage. The situation gets worse if considered that one of the primary application of smart contracts is the handling of valuable assets, including money. It is not uncommon that a bug in a smart contract leads to the loss of large quantities of money~\cite{Vessenes2:2016}. The smart contract community is aware of these risks and tries to counter them with intensive code reviews, including bug bounty programs and the open sourcing of all the code. This processes can be expensive concerning the use of money and time.
\par
To reduce the costs and accessibility related to smart contract development for non-specialized developers with a low budget, a smart contract programming language that is designed to aid the developer in delivering safe and robust smart contracts is advantageous to have. Such a programming language should help to detect and prevent bugs instead of providing pitfalls that are only avoidable by the more experienced smart contract developers. The current programming languages used for programming smart contracts where designed with the primary goal of making it easier for existing developers to transit to the smart contract world. Because of that, their designers tried to use existing concepts and features from non-smart contract programming languages and even copied semantic and syntactic aspects from them. 
\par
It has not been shown that programming in a smart contract environment is similar enough to other programming environments to assume that concepts and paradigms that work there do work similarly well for smart contracts. The potential to use new paradigms and concepts to prevent bugs and attacks like those identified in~\cite{Atzi:2016} gives the motivation for creating a new smart contract programming language which explores this possibility to come up with a solution.

\subsubsection{Motivating Problems}\label{sec:chal}
The challenges and attack vectors identified in~\cite{Atzi:2016} are a core motivation for this work, and it is a primary goal that \textit{Mandala} can prevent these and potentially many other yet undiscovered attacks by design. A summary of some of these challenges, problems and attack vectors is presented to increase the comprehension of this aspect of the motivation and is not meant to be a complete enumeration of all existing problems identified but is meant to include the essential ones which contributed to attacks and bugs resulting in the lost of large sums of money~\cite{Atzi:2016}. The summary uses the Ethereum~\cite{Buterin:2014} blockchain, the Ethereum Virtual Machine(EVM)~\cite{Wood:2015}  and the Solidity~\cite{Solidity:2017} programming language for the examples as they are the currently most used combination.
\newline
\par\textbf{Open Execution Environment}
\newline
Smart contract blockchains like Ethereum are open environments where everybody can deploy code in the form of a smart contract. This code then can interact with other smart contracts. This openness leads to the situation where most of the code that will ever run in the environment is unknown at the time when a smart contract is developed and deployed, and it cannot be assumed that when a message is sent or received that the communication partner is collaborative. The only entity on a blockchain that can enforce a smart contracts behaviour is the virtual machine (current ones do not enforce much). 
\newline
\par\textbf{Reentrancy}
\newline
The EVM is a single threaded virtual machine, but this fact does not mean that a call to a contract is an atomic execution. In the case that a contract calls another contract, this contract can then call back to the contract that called him, which may not expect this behaviour. This is called reentrancy and can create dangerous security flaws if neglected~\cite{Atzi:2016}. A developer has to ensure that whenever another contract is called all invariants hold to prevent reentrancy based flaws. It is not enough to enforce invariants after the execution of a function.  
\newline
\par\textbf{Exception Handling}
\newline
In Solidity, there are multiple situations where an exception can occur~\cite{Atzi:2016}. However, there is limited support for handling these exceptions. Solidity knows two ways of calling a function. One variant returns the result if no error occurred but does not allow to detect an error. The other variant tells the caller if the call succeeded or resulted in an error but in case of success does not allow to retrieve the result of the call. This encourages the developer to ignore errors as they are often more interested in the result than the error.
\newline
\par\textbf{Type Casts}
\newline
Solidity has some statical type checking, but it is not a fully statically typed programming language. One example is type casts on contracts which are neither checked at compile time nor runtime~\cite{Atzi:2016}. If a Solidity function takes an address of a contract of a certain type as a parameter, the actual contract code at that address may be completely different then the code expected based on the type. Due to this a smart contract developer has to treat every call to an address as a call to unknown potential malicious code, except if he formally can prove that it is actually of the expected type.
\newline
\par\textbf{Transfering Ether}
\newline
In Ethereum, there is more than one way to send Ether to a contract. The primary way is to send Ether along with an arbitrary function call. In that case, the receiver can reject the Ether by producing an error or update its internal state to enforce its invariants. Another way is to use the \textit{send} or \textit{transfer} function provided by Solidity, and in that case, the receiver can not update its internal state as the call will not forward enough gas (resource needed to execute operations) with the intention of protecting the caller from reentrancy attacks.
\par
There are two lesser known ways to send Ether where the receiver neither can reject the Ether nor can react on the incoming payment. First, a miner could use a smart contracts address as the coinbase (address to receive the block reward) of a block which then leads to the situation where the smart contract receives the block reward and the transaction fees of the mined block. The second way is to specify a smart contracts address in a selfdestruct operation. If the deleted smart contract does hold any Ether, the specified smart contracts receive the Ether.
\newline
\par\textbf{Contract Selfdestruction}
\newline
The Ethereum virtual machine does allow a contract to issue a so-called selfdestruct which does delete any code associated with it, clears its stored values and transfers all Ether away. This has a negative side effect as any contract that depends on another smart contract must account for the fact that the other can vanish at any time, except if it can be formally proven that the target contract does never execute a selfdestruct.

\section{Methodology}\label{chap:meth}
In the previous section, a brief overview of the \textit{Mandala} smart contract programming language concepts was shown as well as the motivation for it and some problems currently used alternatives have. This Section will show how creating \textit{Mandala} is approached on a methodological level and what the expected outcome is. The project consists of three core parts: A theoretical part consisting of the design and specification of \textit{Mandala}, where a first draft is described in this thesis. The implementation of a toolchain allowing to compile and execute \textit{Mandala}, which can be used to develop smart contracts needed to validate the Hypotheses presented later in this proposal.

\subsection{Toolchain Architecture}\label{sec:arch}
\begin{figure}
\begin{center}
\begin{tikzpicture}
\draw[fill=black!10]  (0,0) rectangle (11,4);    
\draw[fill=red!5] (0,4) rectangle (11,6);    
\draw[fill=yellow!25] (0.5,0.25) rectangle (6,1.75);    
\draw[fill=yellow!25] (0.5,2.25) rectangle (6,3.75);    
\draw[fill=yellow!25] (0.5,4.25) rectangle (6,5.75);    

\draw [->, line width=0.75mm] (3.25,4.25) -- (3.25,3.75);
\draw [->, line width=0.75mm] (3.25,2.25) -- (3.25,1.75);

\draw (8.5,5) node{\large Outside of Consensus};
\draw (8.5,2) node{\large Inside of Consensus};
\draw (3.25,1) node{\large Interpreter};
\draw (3.25,3) node{\large Validator};
\draw (3.25,5) node{\large Mandala Compiler};



\end{tikzpicture}
\end{center}
\caption{Mandala Toolchain Architecture}\label{img:arch}
\end{figure}
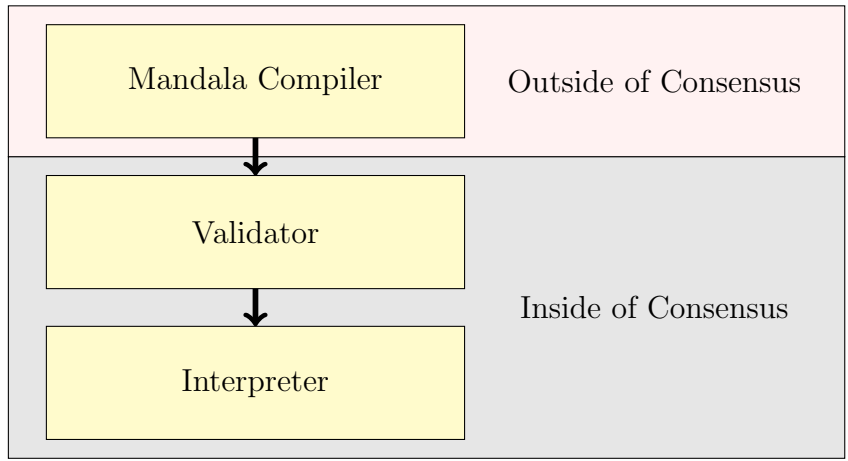

The architecture of the toolchain as seen in Figure \ref{img:arch} is split into two parts the \textit{Mandala} compiler that produces bytecode from the source code and a virtual machine that verifies the integrity of the bytecode and executes it.  The integrity verification has only to be done once when the bytecode is deployed. 
\par
As \textit{Mandala} uses concepts and paradigms that require more guarantees from a runtime then currently established smart contract runtimes provide, existing runtimes cannot be used directly to execute \textit{Mandala}, and as such, the customarily used approach of building a transpiler from \textit{Mandala} bytecode to the existing virtual machines bytecode cannot be used. This means that a new virtual machine for \textit{Mandala} is needed which has to run inside a blockchains consensus to be able to enforce, at runtime, the additional guarantees needed by \textit{Mandala}. To avoid programming a whole blockchain for \textit{Mandala}, to enforce these additional guarantees, this thesis leverages frameworks that allow only to implement the state transition engine of a blockchain and the rest is provided. One such framework is parity substrate~\cite{Substrate:2018} which is part of the polkadot project~\cite{Polkadot:2018}. In case of \textit{Mandala} the state transition engine would consist of the virtual machine that includes at least a validator checking the integrity and an interpreter executing the smart contracts. 
\par
An alternative to this approach would be using a technique presented in SmartDEMAP~\cite{Knecht:2017} where the virtual machine consisting of a verifier and a transpiler are written in another smart contract language and are executed as smart contract on an existing runtime whenever new \textit{Mandala} code is deployed. For this thesis the first approach is used and described in Figure \ref{img:arch} as the second needs to take the limitations of the targeted blockchain into account which would result in a specialised implementation instead of a generalised one.

\subsection{Validation}\label{sec:validity}
For validating the success of the process once the toolchain is in place,  a set of well established and often used smart contract concepts and categories are collected. The range of these may increase considerably while this project progresses. At the current point in time potential candidates would be Fungible and Non-Fungible Tokens, Voting Schemes, Decentralised Autonomous Organisations (DAO) and Initial Coin Offerings (ICO's). These applications are then implemented in \textit{Mandala}. As \textit{Mandala} is more generic and flexible than currently used language, these concepts are wherever possible implemented as reusable components in a standard library. They are implemented by using the concepts introduced over the Hypothesises (See Section \ref{chap:rq} Problem Statement and Research Questions) while avoiding the concepts that they aim to replace. Failing to do so would mean that the used concepts and paradigms are unsuitable for real-world use cases. The resulting programs are compared side by side to existing code in other smart contract language achieving the same functionality to see if \textit{Mandala} can serve as a practical language that delivers on its promises. 
\par
An additional validation step is to collect attacks and problems that happened and are present in other smart contract languages including those shown in the Section \ref{sec:chal} Motivating Problems and analyse them to shows if they can or can not happen in \textit{Mandala}.
\par 
The smart contract application known as Fungible Tokens is used as an example throughout this proposal to show different aspects of \textit{Mandala} and explain its features and differences to other smart contract languages. The most classical example of a fungible token is a basic cryptocurrency where the token contains no other information besides its amount.	
\section{Related Work}\label{chap:relwork}
The most relevant research category to \textit{Mandala} is research related to programming language design and implementation, especially smart contract programming languages. In this work, the term smart contract programming language is used for any programming language that was designed with the intention to develop custom applications that then run inside the consensus process of a blockchain. Classical programming languages (non-smart contract programming languages) are relevant as well as most smart contract programming languages do borrow concepts from established classical programming languages. Some blockchains even use classical programming languages to program their smart contracts even if the language was not explicitly designed with that purpose in mind. Also, classical programming languages can provide guidance and inspirations when designing \textit{Mandala}.

\subsection{Smart Contract Programming Languages}\label{sec:inspandorient}
This section of the related work will cover programming languages that deliver lessons learned and inspirations for \textit{Mandala}. Programming language design is a well-researched topic, and many different programming languages, paradigms and concepts exist already. Not all of the available literature is equally relevant in a smart contract environment. This section will look at various programming languages and inspects how \textit{Mandala} can contribute to the current situation and how it compares to existing smart contract programming languages. 
\par
As smart contracts, geared towards general purpose blockchain programming, are a relatively new concept \cite{Buterin:2014} there are only a few productively used smart-contract programming language of that kind and most others are still in development or count as experimental (see Table ~\ref{tab:langs} for an overview). More narrow smart contract languages existed a while longer science the invention of Bitcoin~\cite{Nakamoto:2008}. All of the smart contract programming languages discussed in this section except for RhoLang~\cite{RChain:2017} have in common, that they were inspired by a classical programming language from which they borrowed parts of their syntax and semantics. There are two reasons for that. First, a smart contract blockchain without a programming language is unusable, and thus the first programming languages had to be developed fast and taking a design that already exists and adapting it for the blockchain is faster than inventing an entirely new programming language with new concepts from the ground up. Second, it makes it easier for existing developers to learn the new programming language as some or even most of the concepts are already familiar to them. The drawback of that approach is that developers may program in the new programming languages similar two how they programmed in the corresponding classical programming language without realising the differences that must be considered~\cite{Hibryda:2016}. Another implication of that approach is that many design decisions are inherited even if they are less beneficial or even harmful in the context of smart contracts and blockchain and better alternatives would have existed.

\begin{table}
  \caption{Blockchain Programming Languages}\label{tab:langs}
\begin{center}
  \begin{tabular}{ | l | l | l | l | }
    \hline
    \textbf{Name} & \textbf{Inspiration} & \textbf{Platform} & \textbf{Remarks} \\ \hline 
    LLL ~\cite{LLL:2017}    & Lisp & EVM  &  \\ \hline
    Serpent ~\cite{Serpent:2017} & Phyton & EVM &  \\ \hline 
    Vyper ~\cite{Viper:2017} & Phyton & EVM & beta \\ \hline
    Mutan ~\cite{Mutan:2014} & Go & EVM & deprecated \\ \hline 
    Solidity ~\cite{Solidity:2017} & Javascript & EVM &  \\ \hline 
    Bamboo ~\cite{Bamboo:2016} & Erlang & EVM & in development \\ \hline
    RhoLang ~\cite{RChain:2017} &  & RChain & in development \\ \hline
    Script ~\cite{Script:2017} & Forth & Bitcoin &  \\
    \hline
  \end{tabular}    
\end{center}
\end{table}
Most of the programming languages as seen in Table ~\ref{tab:langs} are still in development or an experimental state. Of the remaining one, LLL is mainly used as an intermediary programming language when compiling other smart contract programming languages or as a low-level programming language for the EVM~\cite{LLL:2017}. Mutan was deprecated in favour of Solidity. The only left candidates for getting information originating from productive usage are Script, Serpent, and Solidity. Script is very limited in its functionality and was only intended to be used for validating the transfer of bitcoins~\cite{Script:2017}. Solidity is the most used and the feature-richest of the remaining two. Even if not actively used the smart contract programming languages Vyper, Bamboo and RhoLang which lack a productive ready release can give inside in the direction where the field is heading.

\par\textbf{Solidity} is inspired by Javascript which has a focus on providing high productivity, as its primary purpose was to program browser side UI logic, which is typically less security critical than server-side logic and thus the prevention of bugs through language design and delivery of high performance were of secondary concern in its initial design. Over time Javascript's performance has significantly increased to a level where it is even used to develop server applications. Solidity has inherited this focus on productivity in its syntax and semantic but added some minimal additional features to make it more robust against bugs and attacks by introducing a static type system. There are projects like Securify~\cite{Securify:2017} that intend to improve the situation regarding the development of bugfree Solidity by providing formal verification for it. Even if the syntax and semantic is inherited from a productivity-focused programming language, the productivity of Solidity is none the less quite low as many resources have to be invested to do the rigorous auditings and bug hunts needed to ensure that no bugs end up in deployed code were money is at stake.

\par\textbf{Serpent} is inspired by Phyton a currently popular programming language with a focus on delivering high productivity and is often used as a scripting language to automate tasks or analyse data. Serpent, unlike Solidity, has not added a static type system to increase its robustness against bugs and attacks but instead has a smaller feature set compared with Solidity or its inspirer Phyton which makes it more streamlined and eliminates complex features that could accidentally introduce bugs and attack vectors. 

\par\textbf{Vyper} is similar to Serpent inspired by Phyton, it even goes so far, that every valid Vyper program is a valid Phyton program as well. Vyper was designed with the focus of helping the developer to prevent bugs and eliminate attack vectors. Especially the ability of an auditor to quickly discover what a smart contract does is in Vyper's focus. This is done by adding a static type system and aiming for simplicity by removing a lot of Phytons advanced concepts. Thanks to its Phyton syntax and added simplicity Vyper provides compared to Solidity a higher productivity and delivers more robust results.

\par\textbf{Bamboo} is a programming language inspired by Erlang and other Actor based languages. It focuses on explicitly modelling state transitions in contracts and preventing specific attack vectors related to state sharing like reentrancy attacks. Its semantic model is designed with formal verification in mind, guided by lessons learned in the ongoing Solidity formal verification projects. Delivering smart contracts that are more robust is a goal of the project, but besides that, not much is known since the programming language is still in an early stage of development and most information is vague and suspect to change at this time. 

\par\textbf{RhoLang} is not inspired by an existing programming language but instead is build with a process calculi, a fundamentally different mathematical computation model as its core. Process calculi are often used to do formal verification of concurrent or even distributed systems. RhoLang does not only aim to provide formal verification for code written in the programming language but plans to formally verify the compiler and the virtual machine it runs on as well. Thanks to its novel approach it promises a high resistance against attacks and accidental bugs and should be very robust overall. As RhoLang is a deterministic multi-threaded programming language, it has compared to the other languages presented an increased performance. As its programming model is entirely different from what most developers are used to it may have a negative impact on the learning curve. One major drawback of RhoLang is that it needs a new virtual machine and blockchain (RChain \cite{RChain:2018}) as its computation model does not work well with existing smart contract virtual machines.
\newline\par
All these programming languages can be used as inspiration and to extract lessons learned that should not be repeated when designing \textit{Mandala}. The main observation that is of importance for \textit{Mandala} is that all of the described programming languages either add features to make it more robust compared to its inspirer or predecessor or have separate projects providing external tools to make the smart contracts more robust against such flaws. This focus on delivering robust and save smart contracts is especially noticeable in the second wave of smart contract programming languages (the ones currently developed or in experimental use) and most have a shifted viewpoint to make the delivery of smart contracts that are robust against bugs and attacks a central part of the programming language's design. From this observation, it can be concluded, that this property seems to be of high importance for any smart contract programming language. \textit{Mandala} will have the prevention of bugs and attack vectors at the core of its design but will approach this from a different angle.
\par
The current approach in developing smart contract programming languages by orienting them heavily at classical programming languages assumes that smart contract environments are similar enough to other programming environments and implies that proven concepts from classical programming languages are well suited for smart contract programming languages as well. This assumption was never formally researched and has to be considered unproven. Some attacks and problems that are shown in \cite{Atzi:2016} and summarised in the Section \ref{sec:chal} Motivating Problems, do hint in the direction that this assumption may even be harmful concerning the development of smart contract programming languages and applications with few bugs and high attack resistance.
\par
\textit{Mandala} will thus take another approach and evaluate new paradigms and concepts not or rarely used in classical languages and applies them to smart contract languages with the goal of increasing its robustness by using fundamentally different concepts instead of using an existing language and trying to improve it to fit into a smart contract programming environment.

\subsubsection{Comaprison with Mandala}
In this section \textit{Mandala} is compared with the other smart contract programming languages presented in the previous section. The focus in this section is primarily on features, concepts and paradigms that are designed to make the developed contracts less prone to bugs and other attack vectors. This section will ignore RhoLang, as RhoLang has a fundamentally different approach which would make a comparison complicated and mostly meaningless. In comparison to the remaining languages \textit{Mandala} is different in multiple core design aspects, as summarized in Table \ref{tab:langsComp}. The paradigms and concepts mentioned in this sections will be explained in more details later in the Section \ref{chap:mandala} Proposed Solution and Research Idea. This section does focus on the big conceptual differences and does not handle the smaller more nuanced ones.

\begin{table}
  \caption{Blockchain Programming Language Comparison}\label{tab:langsComp}
\begin{center}
  \begin{tabular}{ | l | l | l | l | l | l | }
    \hline
    \textbf{Name} & \textbf{Type System} & \textbf{Method-Dispatch} & \textbf{Isolation} &  \textbf{Access Control}  & \textbf{Decidable}  \\ \hline
    Mandala & Fully Static \footnotemark[1] & Static & Type Based & Capabilities & Yes \\ \hline 
    Serpent & Dynamic & Dynamic & Location Based & ACL\footnotemark[2]  & No  \\ \hline
    Vyper & Weak Static \footnotemark[1] & Dynamic & Location Based & ACL\footnotemark[2] & Yes \\ \hline
    Bamboo & unknown & Dynamic & Location Based & ACL\footnotemark[2] & No  \\ \hline
    Solidity & Weak Static \footnotemark[1] & Dynamic & Location Based & ACL\footnotemark[2] & No  \\ \hline
    \hline
  \end{tabular}    
\end{center}
\end{table}
\footnotetext[1]{Solidity, Vyper and \textit{Mandala} have static type systems, where \textit{Mandala} has a fully static one where no type related errors at runtime can exist. Solidity and Vyper, on the other hand, can have type related problems at runtime.}
\footnotetext[2]{ACL: Access Control List}
The most fundamental differences between \textit{Mandala} and the compared smart contract languages is how they handle state isolation, meaning who can access and modify state. This is of utter importance to get right when providing a programming language that needs to be protected from flaws, bugs and attacks. Solidity and co. are location-based, where code stored under the same address as a persisted storage region has unlimited ability to manipulate that storage region where another piece of code has no access to it at all. In \textit{Mandala}, on the other hand, there is no association between a storage slot (called a cell in \textit{Mandala}) and any code. Instead, only those who can acquire a reference to the cell with the corresponding capability (read and/or write) can interact with the cell. \textit{Mandala} provides opaque types, and references are of this kind and thus references cannot be forged but only retrieved from an entity already able to access it which enables fine granular isolation. 
\par
The unique handling of values together with the strong static type system of \textit{Mandala} is used to enable a different paradigm in respect to how access control is realised. The compared smart contract languages must rely on runtime checks to verify if the caller has the privilege to execute a particular action. The caller is identified by the address of its contract. \textit{Mandala}, on the other hand, does require the caller to provide a value of a specific type, and only if he can obtain or create it, he can execute the guarded function. These values represent capabilities which is a concept originating from security and real-time focused operating systems like EROS~\cite{EROS:2002}. As the type checking happens during compilation \textit{Mandala} can reject programs that attempt unauthorised access at compile-time.
\par
A significant part in making a language better suited to program smart contracts that contain fewer or no bugs and is more resistant against attacks is to make the language accessible to be inspected and audited which enables an auditor to find more problems and increase the possibility that an auditor misses no problem.  \textit{Mandala} has a specific property in this direction that gives it an edge over the other compared languages. \textit{Mandala} is entirely statically dispatched and provides no dynamic dispatches, meaning there is never any uncertainty about what code is executed on a function call.  Another edge of \textit{Mandala} that it shares with Vyper compared to the remaining languages is that both are decidable. Decidability is a concept from the computational theory which states that a program written in that language halts for any possible input. Neither \textit{Mandala} nor Vyper are Turing complete languages which in \textit{Mandalas} case comes with the advantage that an upper bound on the resource usage for every transaction can be computed which means that auditors can ignore the resource management aspects like gas used by the other smart contract languages to prevent non-termination and denial of service attacks.

\subsection{Inspirations from Classical Programming Languages}
Classical high-level programming languages have existed for a much longer time than smart contract programming languages, and the concepts and paradigms used in them have been continuously improved since their invention.  A new smart contract programming language can benefit from the discoveries made during this evolution. When evaluating the concepts and paradigms from classical programming languages in respect to their use in \textit{Mandala}, the difference between classical programming environments and smart contract environments has to be taken into account. Even if it cannot be assumed that something that works for classical programming languages does work for smart contract programming languages equally well, it can still be beneficial to look at those programming languages for inspiration. Different classical programming language concepts and paradigms inspire and influence \textit{Mandala} (See Table ~\ref{tab:langs_gen}, for some core inspiration).
\begin{table}
  \caption{Inspiring Classical Programming Languages}\label{tab:langs_gen}
\begin{center}
  \begin{tabular}{ | l | l | }
    \hline
    \textbf{Name} & \textbf{Language Concepts and Paradigms}  \\ \hline 
    Rust~\cite{Rust:2017} & Ownership \& Borrowing \\ \hline
    Go~\cite{Go:2017} & Composition over Inheritance  \\ \hline
    Java~\cite{Java:2017} &  Class Loader, Class File Verifier \\ \hline
    Haskell~\cite{Haskell:2017} & Algebraic Datatypes, Immutability \\ \hline 
    SML~\cite{SML:1997} & Algebraic Datatypes, Immutability, Modules, Cells, Opaque Ascriptions \\
    \hline
  \end{tabular}    
\end{center}
\end{table}
Smart contract virtual machines, especially the Ethereum virtual machine \cite{Wood:2015} are relatively low-level and often behave more like a CPU than other modern virtual machines. Lately, there was an increase in programming languages that compile to native code instead of bytecode for a virtual machine and still being able to keep most of the concepts provided by modern high-level programming languages. Go, and Rust does fall into this category. These programming languages deliver inspiration for theories and design philosophies that do not need a feature-rich VM to be executed efficiently. On the other side, there are advanced virtual machines like the JVM~\cite{JVM:2017} which can give inspiration on how a good runtime environment can be designed.	
\section{Problem Statement and Research Questions}\label{chap:rq}
The discussion on shortcomings above result in the problem statement for this thesis as follows:
\par\textit{The usage of smart contract programming languages based on concepts and paradigms invented for less safety and security sensitive programming environments than a smart contract blockchain repeatedly led and will lead to bugs and vulnerabilities that can cause a considerable financial loss.}
\par
To address the problem statement and find a way to improve the situation, the approach that is taken in current smart contract programming language design and implementation is questioned, and two research questions are formulated that ask for a different approach than the one currently used. These research questions will guide the development of \textit{Mandala} and influence the decisions made in \textit{Mandala}'s design and implementation.
\par\textit{Research Question 1: What compilation and execution techniques can enforce compile-time guarantees that enable the use of new concepts and paradigms in an open and adversarial execution environment?}
\par\textit{Research Question 2: What programming language paradigms and concepts can improve the robustness against bugs and attacks of smart contracts compared to the currently used ones?}
\par
To explore the research questions two hypothesises are presented that explore two alternative paradigms that provide a potential path for finding answers to the proposed questions and improving the situation described in the problem statement.
\subsection{Hypothesis 1}
\textit{Using opaque and substructural types in a smart contract language allows to use alternative location independent isolation mechanisms that increase the robustness against bugs and attacks in an open and adversarial execution environment}
\par 
Opaque types is a concept used in functional languages like ML (opaque signature matching)~\cite{SML:1997} where only code contained in the same module as a type declaration is aware of the inner workings and how to read and create values of a type. Substructural types~\cite{Walker:2002} is a concept partially (in the form of Unique Types) used in languages like Rust which gives guarantees of how often a value of such a type is used. In classical programming languages, this is used to improve memory and stack optimisations especially in respect to garbage collection and memory shared between threads. The Hypothesis claims that these two concepts together can not only be used for formal verifications and optimisation but can provide an alternative isolation mechanism where the isolation happens on a per type/value basis instead of a per contract basis. This promises a new programming paradigm better suited for smart contracts then the current approaches based on storage isolation.

\subsection{Hypothesis 2}
\textit{Using a capability-based approach to access control in contrast to the currently used access control list based approach in smart contract languages allows to reason about access control at compiletime and provides a more flexible mechanism to safeguard interactions between different smart contract applications}
\par 
Currently used smart contract languages and execution environments like Solidity, use access control list based safeguarding mechanism. On a function call, they identify the caller and check if he has access by looking up his address in a map recording the access rights. These maps can be statically filled with addresses or dynamically managed at runtime. One standard approach is to capture the creator of a smart contract and give him additional rights sometimes including the right to provide access rights to other or even transfer his rights. Capability-based systems originate from operating systems, especially security and reliability-focused operating systems like EROS~\cite{EROS:2002}.  A Capability is a reference to a resource that simultaneously encodes the access rights to that resource. The Owner of a capability has the right to permanently or temporarily give the capability to another process including all or a subset of the access rights of the original.
The Hypothesis claims that a capability-based approach allows capturing most access control related bugs and attack vectors already at compiletime and that this mechanism is better suited for smart contracts. Further, this approach reduces the runtime overhead of access management and may even eliminate it in some cases reducing the expensive store operations needed for managing them in languages like Solidity.
\section{Proposed Solution and Research Idea}\label{chap:mandala}
The problem statement and research questions provided guidance on what to address when designing \textit{Mandala} while the Hypotheses, promise a solution. This section will showcase ideas and concept, based on the Hypotheses, which are intended to represent the foundational aspects of \textit{Mandala}. A preliminary design for \textit{Mandala} will be presented with the help of code examples which showcase a Token like for example a new cryptocurrency. Also, a Purse that allows everybody to deposit such Tokens in it, but only the owner of the Purse to withdraw them is introduced.

\subsection{Mandala Core Design Philosophy}\label{sec:mandalaCorePhil}
\textit{Mandala}'s core goal is not to replace smart contract programming languages like Solidity or Vyper. The existing EVM based programming languages try to give the developer access to the underlying blockchain technology and expressively do that by giving the developer as many tools and freedoms to solve a problem as possible. As already mentioned, this approach does have the drawback, that it is easier to introduce bugs leading to unexpected behaviour and that it is harder for an auditor to spot these bugs in a smart contract~\cite{Hibryda:2016}. 

\textit{Mandala}'s core design goal is to reduce the possibilities of unexpected behaviour that is introduced by accident and make it easy for an auditor to reason about the code. \textit{Mandala} should still be as expressive as possible and covers any necessary features and concepts to be usable as a practical smart contract language. In case of a trade-off between expressiveness and safety or auditability \textit{Mandala} will prefer the approach of increasing safety or auditability. In the following, two qualitative attributes are defined, which are used to evaluate the robustness against bugs and attacks with an explanation of what they mean in the context of \textit{Mandala}. 
\par\textbf{Safety} means that the programming language is designed in a way that prevents specific exploits and makes it easy to write safe code and hard to introduce exploitable code without intention. The obvious way to code something in \textit{Mandala} should always be the safe way to code it. \textit{Mandala} should eliminate the risk that a developer can develop code that is vulnerable to specific well-known exploits, like the one presented in the Section \ref{sec:chal} Motivating Problems. Further \textit{Mandala} will select its feature set primarily based on safety aspects and only secondary on expressiveness and performance aspects. 
\par\textbf{Auditability} means that it is easy to understand what a piece of code does and conclude how it would behave if executed. \textit{Mandala} will focus on local code, meaning that an auditor does not need to know the whole program to reason about an individual piece of code. If context information is needed to audit a piece of code, it should always be clear where to look for it, and it should be unambiguous how it influences the currently audited code. Beside manual auditing, the auditability includes the ability of \textit{Mandala} code to be analysed by other programs, like automatic bug finders or formal verifiers. 

\subsection{Mandala Language Design by Example}
This section will present Mandala with the help of code examples followed by some descriptions of the concepts and feature used in the corresponding Examples. This is not meant to cover all features that \textit{Mandala} should have in the end.

\subsubsection{Token}\label{sec:Token}
The Token module in Listing \ref{lst:Token} provides a type that can be used to represent tokens including the necessary functions needed to split, merge and create those tokens.

\begin{lstlisting}[language=mandala, label=lst:Token, caption=Token Module]
module Token {
  
  type Drop Persist Token[T](UInt)                

  risk NumericOverflow
  public merge[T](Token[T](amount1), Token[T](amount2)) => {
    Token[T](amount1 + amount2)                                       
  }    

  risk NumericUnderflow
  public split[T](Token[T](amount), split:UInt) => {
    (Token[T](amount-split), Token[T](split)) 
  }
  
  protected[T] mint[T](amount:UInt) => Token[T](amount) 
  public default[Token] zero[T]() => Token[T](0)

}  
\end{lstlisting}

\par\textbf{Modules}
\newline
Line 1 in Listing \ref{lst:Token} does declare a module which encapsulates components (types, capabilities and functions)  between the curly braces. Modules are \textit{Mandalas} way to group related code together and it is immutable after deployment, meaning that no components can be added, modified or removed from a module once its content is declared.  Additionally, components defined in the same module have extended rights regarding each other and can do specific actions that components from other modules can not. 
\newline
\par\textbf{Types}
\newline
On line 3 in Listing \ref{lst:Token} a type component is declared. \textit{Mandala} uses algebraic data types (ADT) as its core value representation. An ADT is a type that has multiple constructors, and each of the constructors can have multiple fields. When a value is created over a constructor, the values supplied for the fields are stored in the value and can later be accessed by unpacking the value. A value can be unpacked by providing a piece of code for each constructor, and at runtime, only the code corresponding to the constructor used to create the value is executed and gets access to the parameter used to construct the value. The type Token in the example does only have one constructor, which takes an unsigned integer as field parameter. The type declaration on line 3 in Listing \ref{lst:Token} is generic, meaning it can be parameterised over another type denoted T in the example. This means that the declared type does not represent a single type but a full type family. For Example, Token[Eth] and Token[Btc] are two different types which belong to the same family. An ADT value can only be created by the module defining the type unless the ADT is marked with open in that case everybody can create new instances.
\newline
\par\textbf{Capabilities}
\newline
By default, all values are restricted in how functions can interact with them. The only thing they can do with a value is passing it to another function or using it as an argument to create another ADT. Even accessing its fields, making a copy of it or throwing it away is forbidden. To allow further operations, Mandala provides so-called capabilities which can be attached to values and are tracked statically by the type system.  A capability can only be attached by the module defining the capability unless the capability is marked with open in that case the module defining the type to which the capability is attached to, can do so as well. Detaching a capability, on the other hand, can be done by everybody. Two values that have the same base type but different capabilities are treated as different types by the type system. Beside predefined capabilities, \textit{Mandala} supports custom capabilities (See line 3 in Listing \ref{lst:Purse}), which can be used as an access control mechanism to protect access to components in a module. The type on line 3 in Listing \ref{lst:Token} for example has per default the capabilities Drop and Persist. This allows everybody to drop the value without using it (Drop) and enables the value to be persisted (Persist) (see Section \ref{sec:Purse}), but as still nobody can make copies of the value and only the defining module can create new values it is well suited to represent a token or other asset. 
\newline
\par\textbf{Functions}
\newline
Beside types, a \textit{Mandala} module can contain functions which are similar to functions from well-established languages. To make interaction with ADTs easier, they can be unpacked at the place where they are received as a parameter (See lines 6 and 11 in Listing \ref{lst:Token}). Like types, functions can have generic type parameters which allow defining a function once for a whole family of types. In Mandala functions cannot be recursive and further do only support static dispatch, meaning that it is known during compilation what code will be executed when a function call is executed. The enforcement of non-recursivity is achieved by allowing only function calls to already deployed functions requiring that functions in the same module be deployed one after another. The function on line 16 in Listing \ref{lst:Token} is marked with default, which tells the \textit{Mandala} compiler that it should use this function when a default value for the Token type is needed (see Section \ref{sec:Purse}).
Every function has a visibility which defines who can call that function. The functions on the lines 6, 11 and 16 in Listing \ref{lst:Purse} are public and can be called by anyone. A private function could only be called from the same module. The mint function on line 15 in Listing \ref{lst:Token} has the protected visibility which is linked to the functions generic type parameter T. Code can only call a protected function if its module defines the type to which the protected visibility is linked. The mint function, for example, can only be used to mint tokens by the module defining the token's type. As an example, a Token[Eth] can only be minted by the Module defining the Eth type. 
\newline
\par\textbf{Error Handling}
\newline
The presented merge and split functions in Listing \ref{lst:Token} on the lines 6 and 11 allow to take a Token and split it in two or take two Tokens and merge them. These two functions ensure that the balance of their incoming and outgoing tokens sum up to the same amount.  \textit{Mandala} uses save arithmetic, and thus an underflow or overflow error can happen which is represented with the risk declaration on the lines 5 and 10 in Listing \ref{lst:Token}. When an error occurs, the progress made in the function is rolled back. The caller of the function receives the initial arguments to the function (to preserve non-copyable arguments) together with an error code as the return value. The caller can decide to either handle the error or make a rollback itself and forward the error.

\subsubsection{Purse}\label{sec:Purse}
The Purse module in Listing \ref{lst:Purse} provides a type that can be used to deposit and withdraw tokens from it similar to a bank account, including the necessary access control mechanisms to keep the funds safe.

\begin{lstlisting}[language=mandala, label=lst:Purse, caption=Purse Module]
import Token.*
module Purse {
  open capability Withdraw 

  open type Persist Withdraw Copy Drop Purse[T](
    Persist Copy Drop Modify Ref[Persist Drop Token[T]]
  )

  risk NumericOverflow
  public active deposit[T](Purse[T](tokenRef), deposit:Token[T]) => {
    modify tokenRef with Token(t) => merge(t, deposit)
  }

  risk NumericUnderflow
  public active withdraw[T](Withdraw Purse[T](tokenRef), amount:Int) => {
    modify tokenRef with 
        Token(t) => case split(t,amount) of 
            (rem,split) => rem & return split
  } 
}
\end{lstlisting}
\par\textbf{Imports}
\newline
Line 1 in Listing \ref{lst:Purse} shows how functions and types from modules can be imported, such as they can be used in another module. The example code imports the Token module which is colocated in the same namespace as the Purse module. In the end \textit{Mandala} will provide a more elaborated namespace and import system able to handle a large number of modules. For simplicity, the remaining examples will skip the import and assume that all necessary components are imported.
\newline
\par\textbf{Cells and References}
\newline
Line 5 in Listing \ref{lst:Purse} is a type declaration similar to the Token from the previous example. The most significant difference is that it uses an argument of type Modify Ref[Token[T]] for its sole constructor. Values of type Ref[X] represent a reference to a cell persisting a value of type X. There can be multiple references to the same cell allowing to share the stored value. The Modify capability allows using the reference to exchange the value stored in the cell with another value. References without that capability could only read the value. A cell can only store the value of a type with the Persist privilege. Everybody can generate a cell, but it initially would be empty or filled with the default value if one is defined as it is the case for Tokens  (see Section \ref{sec:Token}). The creator of the new cell does only get a reference to the cell and not the cell itself.  
\newline
\par\textbf{Effect System}
\newline
The functions on line 10 and 15 in Listing \ref{lst:Purse} have a new keyword called active. By default, any function is pure, which means that given the same arguments it will always return the same result and does not have any side effect. As these two functions modify a cell, they have a side effect and thus must be marked active. Beside active, there are the modifiers init which is similar to pure, but the function is allowed to create new cells. Functions marked with dependent can additionally read values from cells, and active ones can even write values to cells. This is an effect system which makes it easier for an auditor to see what is going on and it further can prevent errors. The modify with expressions on the lines 11, 16 in Listing \ref{lst:Purse} for example does take a pure expression that takes the current cells content and returns a new value to be written back into the cell. As this expression is pure, it is guaranteed that it does not interact with other cells while the value is under modification. This prevents shared state-based attacks.

\subsubsection{Purse Storage}\label{sec:Store}
The PurseStorage module in Listing \ref{lst:Store} provides functionality that allows everybody to retrieve the Purse associated with an individual identity and deposit Tokens in it or even withdraw Tokens if he or she is the owner of the Purse.

\begin{lstlisting}[language=mandala, label=lst:Store, caption=Purse Store Module]
module PurseStorage{
  open type Store[T](Copy Context[Token[T]])

  public getMyPurse[T](id:Master ID, Store[T](c)) => Purse(derive(c,id))
  public getPurse[T](id:ID, Store[T](c)) => Purse(derive(c,id)).detach[Withdraw] 
  
  risk NumericOverflow
  risk NumericUnderflow
  public active transfer[T](src:Master ID, to:ID, store:Store[T], value:Int) => {
    deposit(
       getPurse[T](to,store), 
       withdraw(getMyPurse[T](src,store),  value)
    ) 
  }
}
\end{lstlisting}
\par\textbf{Identification}
\newline
In Listing \ref{lst:Store} on line 4, 5 and 9 a new type called ID and a capability called Master is used. The ID type is a primitive type similar to the address type used in solidity and the ethereum virtual machine. Everybody can generate an ID if he knows the corresponding identification string. However, an ID with the Master capability cannot generally be generated. Every keypair used to access the blockchain running \textit{Mandala} is associated with an ID. To obtain a value of the type Master ID that can be used in a transaction the transaction has to be signed with the associated private key. Besides using a private key, Master ID's can be created by calling the new function in the ID module which then produces a unique ID with the Master capability. The new function does guarantee that it will never produce the same ID twice and that each ID is different from all ID's associated with private keys.
\newline
\par\textbf{Contexts}
\newline
In Listing \ref{lst:Store} on line 2 a type called Context is introduced. A Context is similar to an ID as new unique once can be created by calling a new function in the Context module. Unlike ID's, Contexts have a generic type parameter and as such their exist a whole family of Context types, one for each existing type. The primary purposes of Contexts are to associate an ID with a reference. Given an ID and a Context[T] someone can call the derive function (see line 4 and 5 in Listing \ref{lst:Store}) to generate a Reference to a Cell containing a value of type T. Using the same ID and Context as derive input always results in the same reference while using a different combination always yields a different reference. A Context can be seen as a storage area for cells where each cell is associated with an ID. From this viewpoint, the Store type provides a 1 to 1 association between IDs and Purses.

\subsubsection{Token Instantiation}\label{sec:Inst}
The MyFixSupplyToken module in Listing \ref{lst:Instance} declares a new Token with the help of the previously presented modules and deposits the initially minted Tokens into the deployers Purse.

\begin{lstlisting}[language=mandala, label=lst:Instance, caption=Token Instance Module]
module MyFixSupplyToken {
    public type MyToken

    public val defaultStore = Store[MyToken](Context.new[Token[MyToken]]())

    risk NumericOverflow
    init(deployer:Master ID) => deposit(
       getMyPurse(deployer,defaultStore), 
       mint[MyToken](100000000)
    )  
    
}

\end{lstlisting}

\par\textbf{Constant Values}
\newline
For \textit{Mandala} to be useful, there has to be a way to store a value globally in a way that it is accessible without the need to possess already another value (like it is the case with cells and references). Depending on the blockchain model in which \textit{Mandala} is used this could be delegated to another layer. For example, a UTXO based blockchain could store \textit{Mandala} values in UTXO's and an account based blockchain could store them in the accounts. To be independent to a specific blockchain model line 4 in Listing \ref{lst:Instance} shows an alternative where \textit{Mandala} provides a top-level storage slot called val. A val represents a constant that is initialised when the contract is deployed and does never change afterwards. To be independent of the point in time when the module is deployed the vals initialisation expression must be of pure or init effect. As a val can be used more than once over the span of multiple transactions its content must have the Copy and Persist capability.

\par\textbf{Initialisation}
\newline
In Listing \ref{lst:Instance} on line 7 an initialisation function that has precisely one Master ID parameters is provided. This function is executed exactly once when the module is deployed. The received Master ID is the one associated with the deployer of the module. This allows a hook to initialisation the module. If the initialisation produces an error, then the module will not be deployed. In Listing \ref{lst:Instance} on Line 8 to 9, for example, a fixed amount of Tokens is created and put into the Purse of the deployer.

\subsection{Discussion of the Mandala Language Proposal}
The two core qualitative goals of Mandala concerned safety and auditability. This section will show how the presented approach for Mandala aims to achieve these goals.

\subsubsection{Safety}
Safety is concerned with the elimination of the risk for a developer to introduce bugs, flaws and other problems into the code without the intention to do this. The Section \ref{sec:chal} Motivating Problems provided examples for such problems that are a concern in other smart contract programming languages. This section will look at these problems and show how Mandala can address them.
\newline
\par\textbf{Open Execution Environment}
\newline
By default \textit{Mandala} values have no capabilities, and nearly nothing can be done with them. The defining module can still do more with the value as it is entitled to attach open capabilities or capabilities defined in the same module. This allows declaring value types that enforce guarantees even against code not yet deployed. Further \textit{Mandala} uses such values as capabilities to control who can do what with it. This allows enforcing that a module/function needs to be handed a capability willingly before it can interact with the protected resources. This allows a programming style where no manual protection layer that specifies what has to be done to acquire access to a resource has to be provided. This has the advantage that nothing has to be known about other potentially later deployed code to be protected from it as all code plays by the same rules independent of when and where and by whom it is deployed.
\newline
\par\textbf{Reentrancy}
\newline
Reentrancy problems occur when a state is shared between invocation of functions that do not enforce invariants in its intermediary state which then is leveraged by an attacker. The classical reentrancy attack where the same function or a function in the same contract (module in case of \textit{Mandala}) is called cannot happen at all in \textit{Mandala} as \textit{Mandala} does only support static dispatch and does not allow recursive calls, or other circular dependencies.
\par
As Mandala cells are accessed over a reference, multiple Modules could obtain a reference and gain access to the same shared state. To prevent inconsistent invariants of a cell during its modification \textit{Mandala} provides a particular mechanism to modify a cell where the modification is represented as a pure state transition function. This guarantees that when a cell is modified in this way no other function can access any other cell while a cell is changed. A developer can misuse the modify mechanism to reopen itself to these kinds of attacks but it would require an effort to do so and would be complicated and sophisticated and would be easy for an auditor to spot and investigate.
\newline
\par\textbf{Exception Handling}
\newline
Mandala does not inherit the weaknesses of Solidity and other EVM based languages in respect to error handling because errors are not communicated over the classical return path, and thus a caller can do both handle the error and consume the return value on success. Mandala further enforces that all potential errors are documented in the function signatures, and thus the caller does always either need to handle the error case or explicitly declare the error again delegating the error handling to its caller.
\newline
\par\textbf{Type Casts}
\newline
Mandala has a type system that does not have the concept of a typecast as any value is of exactly one type. It could be argued that attaching and detaching capabilities is a typecast, but these are checked at compiletime and cannot be misused to execute unexpected code like it is possible in solidity. \textit{Mandalas} type system ensures that all type related errors are detected during compilation and no type related errors can occur at runtime.
\newline
\par\textbf{Transfering Ether}
\newline
\textit{Mandala} will provide a completely independent definition of Ether (resp the native cryptocurrency of the targeted blockchain), realized over the Token framework from the Listings \ref{sec:Token}, \ref{sec:Purse} and \ref{sec:Store}. Moreover, as such, it does not gain any special treatment, and the only way to transfer Ether is to call a function that takes a Token[Eth] as a parameter. This concept guarantees that the receiver is always equipped to handle the Ether and can react appropriately. 
\newline
\par\textbf{Contract Selfdestruction}
\newline
\textit{Mandala} does not know the concept of a contract. \textit{Mandalas} module is the nearest thing to a contract. Modules can be created but never destroyed preventing any self-destruct related problems as this concept is not known. If cells are considered as the state part of a contract, then the cell either has a default value, or the compiler will enforce the developer to specifies the behaviour in case the cell is empty. 

\subsubsection{Auditability}
\textit{Mandala} has certain core aspects that make the job of an auditor easier. The first is that \textit{Mandala} does not have dynamic method dispatch and does not support recursion even indirect ones. This means that an auditor can know for sure what code is executed by a function call and even the deployment of other code in the future cannot change that. Second, the non-recursivity ensures to an auditor that all code executes and terminate without having to consider unexpected events like out of gas exceptions (common in EVM based language). This is the case because \textit{Mandala} allows enforcing upfront (before the transaction is executed) that enough resources are provided for any possible execution path. Moreover, \textit{Mandalas} strong static type system with its restricted types tells an auditor precisely what can be done with values of a type without needing him to inspect any code that interacts with the value. This allows to analyse a module and check its integrity even if the auditor does not know how other code will use the module and its types. Lastly, the improved exception handling system (compared to Solidity and co.) does make it easier for an auditor to check that every corner case is handled correctly since exceptional cases have to be declared in the function signature. As many problems are related to unwanted state changes, Mandalas effect system can indicate to an auditor where to look for these problems and what functions can be safely ignored when investigating state-related problems. 			
\section{Summary and Preliminary Conclusions}
This research proposal analysed the current situation for existing smart contract programming languages. The proposal identified the need for improvement in safety and auditability aspects. Then the thesis provides research questions and hypotheses that promise a way to develop a more robust, safer and easier to audit smart contract programming language that addresses the existing problems and challenges by using alternative concepts and paradigms.
\par
Guided by the path presented by the research questions and hypotheses a new smart contract programming language called \textit{Mandala} is proposed. \textit{Mandala}'s overall design goals focus on improving safety and audit-ability while retaining the practical usability and promises to prevent currently existing weaknesses and attack vectors that plague current smart contract programming languages. 
\par
\textit{Mandala} provides concepts like substructural and opaque types that allow expressing concepts like cryptocurrency tokens and other assets directly as first-class citizens of the programming language. These types are realised and extended over so-called \textit{capabilities} an idea inspired by security-oriented operating systems, which allows enforcing access to protected functionality already during compiletime instead of runtime. 
\par
The research proposal shows that in theory alternative programming language concepts and paradigms that leverage compilation in a trusted execution environment can provide a viable solution to improve the current state of smart contract programming.

\end{document}